# Organ Dose Conversion Factors for Murine Galactic Cosmic Ray Irradiation

Authors: S. Hosseini[1], M. Sivertz[2], E.M. Alves[1], L.M. Carter[3*], and M.D. Story[1*]

Affiliations:
[1]Department of Radiation Oncology, University of Texas Southwestern Medical Center, Dallas, TX;
[2]NASA Space Radiation Laboratory, Brookhaven National Laboratory, Upton, NY;
[3]Department of Medical Physics, Memorial Sloan Kettering Cancer Center, New York, NY;

[*]Corresponding authors

Current addresses:

SH: Huntsman Cancer Institute, University of Utah Health, Salt Lake City, Utah
EMA: Rush University, Department of Radiation Oncology, Chicago, Illinois 60612
LMC: Memorial Sloan Kettering Cancer Center, 1250 1st Ave., New York, New York 10065
MDS: Mayo Clinic Florida, 4500 San Pablo Rd South, Jacksonville, Florida 32224

## 1 ABSTRACT

Galactic cosmic rays (GCR) are a principal source of ionizing radiation exposure for astronauts during deep space missions. Given the ambition to expand manned space exploration to distant destinations like Mars, it is essential to accurately predict the radiation doses astronauts are likely to encounter and the consequent biological impacts. Accurate dose predictions are important for operational radiation safety, ensuring that risk assessments and protective measures are appropriately calibrated to the myriad of challenges of deep space travel.

The GCRsim facility at the NASA Space Radiation Laboratory enables small animal radiobiology studies of GCR exposure, offering a controlled setting to mimic the complex radiation conditions found in deep space. This manuscript introduces a series of Dose Conversion Factors (DCFs) which enable rigorous absorbed dose calculations for mice irradiated at the GCRsim.

METHODS: A formalism was introduced for calculating organ-level and voxel-level radiation dose to a representative mouse phantom, based on DCFs quantifying radiation absorbed dose per unit fluence of different GCRsim species for different irradiation orientations. The PHITS Monte Carlo code was employed to compute the DCFs in units of $Gy \cdot m^2 \cdot ion^{-1}$.

RESULTS: A library of murine DCFs were derived using the PHITS Monte Carlo code for six irradiation orientations: right-left, anterior-posterior, superior-inferior, and their opposed variations. Absorbed doses to the murine total body were calculated with the method and compared with ion chamber measurements, which agreed within 10%.

CONCLUSION: A library of dose conversion factors for mouse irradiation at GCRsim was developed and validated against physical measurements. These DCFs account for organ-specific variations in radiation dose from different GCR species., enabling improved assessments of potential radiogenic effects, toward improving astronaut safety measures for future deep space missions.

## 2 INTRODUCTION

Interplanetary spaceflight presents unique challenges and risks, among which radiation exposure remains a critical concern for the safety and health of astronauts. Exposure to space radiation increases risk for various health detriments including central nervous system damage, cardiovascular disease, and carcinogenesis (*1*). The space radiation environment is a complex and dynamic blend of high-energy particles originating from sources inside and beyond the galaxy; beyond low Earth orbit, the conditions significantly diverge from the terrestrial radiation field, particularly in aspects like radiation quality and energy levels. Major sources of astronaut exposure in deep space include solar particle events and Galactic Cosmic Rays (GCR). Unlike the more sporadic solar particle events, GCRs provide a steady exposure to radiation that varies minimally with the solar cycle, thereby posing a constant challenge for radiation exposure management during long-duration missions beyond Earth's protective magnetosphere.

GCRs, originating from outside our solar system, are composed of a high-energy spectrum of fully ionized nuclei spanning the periodic table; however, the dosimetrically dominant species for space travel applications are considered to include hydrogen (H), helium (He), and "HZE" ions of carbon (C), oxygen (O), neon (Ne), silicon (Si), calcium (Ca), and iron (Fe) (*2*). The high energy and penetrating power of these particles mean that they can traverse through shielding materials and human tissue, leading to direct ionization and secondary particle cascades.

Understanding interactions between GCRs and biological systems is fundamental to estimating the radiation doses astronauts receive and for assessing consequential bioeffects. The *Galactic Cosmic Ray Simulation* (GCRsim) at the NASA Space Radiation Laboratory (NSRL) at Brookhaven National Laboratory offers a sophisticated empirical means of simulating the heterogeneous GCR environment encountered in deep space. This facility utilizes heavy-ion beams extracted from Brookhaven's Booster accelerator and allows for the rapid switching of ion species and energies, closely mimicking the operational GCR spectrum. GCRsim can be configured for live-animal irradiations, such as with rodents, providing a crucial ethical alternative for collecting data that cannot be gathered from human subjects. By accurately mimicking the GCR spectrum, controlled experiments can be conducted that yield insights into GCR-specific biological effects (e.g., mixed-field, HZE effects), thereby improving the predictive accuracy of radiation bioeffect and risk models for astronauts on long-duration deep space missions.

The GCRsim facility allows researchers to specify a nominal radiation absorbed dose for irradiations; however, this nominal dose is calibrated to ion chamber measurements in a reference geometry and can introduce systematic errors when applied to subjects with non-standard geometries, such as mice. Furthermore, variations in tissue density, geometry, and the beam orientation, which may significantly influence the absorbed doses received by individual organs, are not accounted for.

In this study, we derive a library of dose conversion factors (DCFs) for mouse irradiation with the GCRsim using different beam orientations. We evaluate the dose calculations against physical dosimetry measurements. The DCFs enable absorbed doses delivered by the different GCR species to be rigorously accounted for, and allow organ-level and voxel-level dose reconstruction from measured fluences of each beam component or from the nominal or physical dose measured via ion chamber. Moreover, the derived dose conversion factors allow for the assessment of equivalent doses by incorporating the linear energy transfer (LET) characteristics of the different species. Ultimately, the DCF library enables more rigorous analyses of potential dose-response relationships in organs of mice irradiated using GCRsim, in turn enabling better insights into potential radiogenic cancer induction and tissue reactions, which is imperative for developing and optimizing protection strategies and countermeasures to ensure astronaut health on missions to Mars and beyond.

# 3 MATERIALS AND METHODS

## 3.1 THE SIMULATED SPACE RADIATION ENVIRONMENT

The NSRL's GCR simulator facilitates ground-based radiologic experiments for space applications (*2*). Prior to the development of GCRsim, most ground-based experiments used only one type of ion beam at a specific energy. In contrast, GCRsim can handle multiple beams with varied energies and switch between them in a short time (*3,4*). This allows for an effective simulation of complex multi-ion irradiation scenarios with different energy levels, LETs, and dose rates. Two configurations of the GCR mixed ion beam are available: 1) the full GCR simulation characterized by the use of seven ion species at various energies, and 2) the simplified GCR simulation (SimGCRsim) which involves 5 different ions including hydrogen at 2 different energies (Table 1).

**Table 1:** Ions, energies, and fractional dose contributions of NSRL simplified 5-ion beam (SimGCRsim).

| Ion | Energy (MeV/n) | Dose Fraction |
| --- | --- | --- |
| H | 1000 | 35% |
| Si | 600 | 1% |
| He | 250 | 18% |
| O | 350 | 6% |
| Fe | 350 | 1% |
| H | 250 | 39% |

In this study, simulations of organ-level absorbed dose and absorbed dose distribution in a rodent model were evaluated using whole-body dosimetry instrumentation and counting systems associated with the GCRsim beamline.

## 3.2 NSRL BEAM CALIBRATION

### 3.2.1 Beam Preparation

Prior to calibration, the NSRL beam is magnetically shaped to form an approximately square spot with a uniformly intense target region. Uniformity is assessed by imaging the beam spot with either a CCD camera or a large pixelated ion chamber, yielding a root-mean-square intensity variation across target-region pixels typically below 3%.

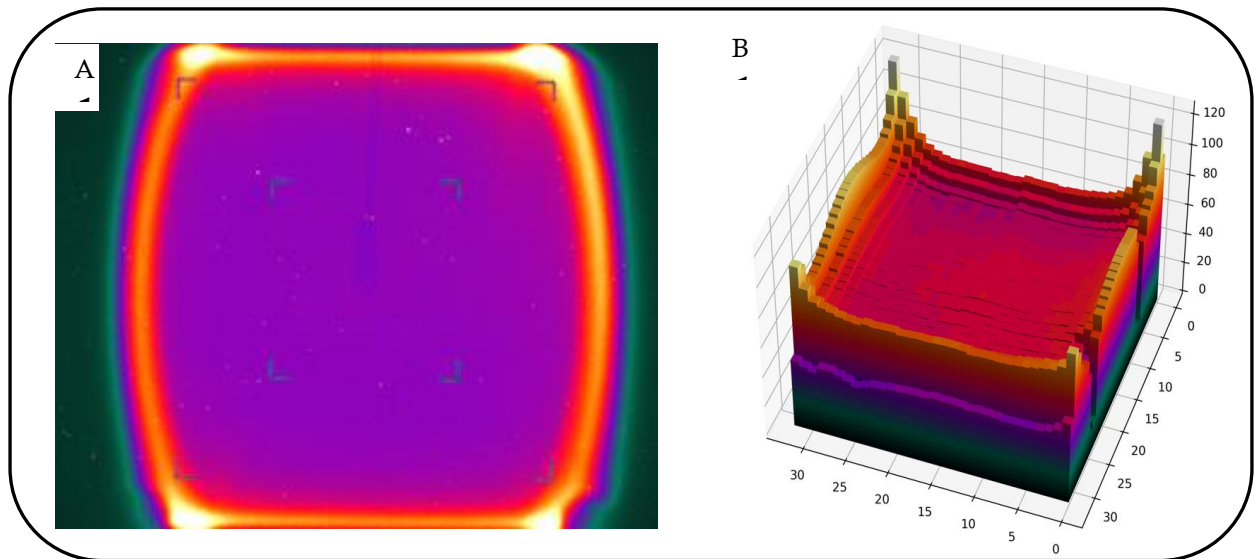

**Figure 1:** Beam profile captured using the Digital Beam Image Camera (A), and the 1024-pixel ion chamber (B).

### 3.2.2 Beam Energy Measurement

For sufficiently low beam energies, the Bragg curve is measured by inserting high-density polyethylene (HDPE) into the beam path and recording the energy deposited in ion chambers positioned upstream and downstream of the HDPE degrader. The ratio of downstream to upstream ion chamber responses is plotted to identify the Bragg peak, from which the beam energy is determined.

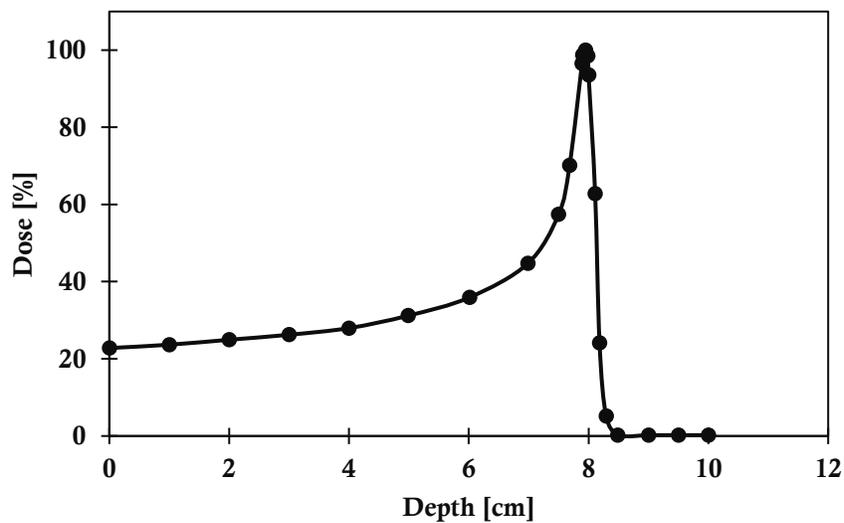

**Figure 2:** Bragg curve for 103.9 MeV protons in HDPE.

### 3.2.3 Dose Calibration

Once the beam energy has been measured and the beam spot tuned for good intensity uniformity, a small thimble ion chamber is installed at the sample location in the center of the beam spot. The ion chamber is a hollow sphere made of tissue-equivalent plastic with an internal volume of 1 cc of air at ambient temperature and pressure. A high-voltage electrode extends into the center of the sphere to collect the charge produced when ions pass through the air, and is maintained at –300 V. Periodically, the ion chamber is sent to a NIST-traceable laboratory to check the calibration against a $^{137}$Cs gamma source. The calibration lab provides a conversion factor in roentgen per coulomb, relating the charge produced in the ion chamber to the radiation exposure level. Conversion from the exposure to dose-in-air is performed using the Bethe–Bloch formula, with corrections for electron shell effects, nuclear interactions, and low-energy neutralization effects. The calibrated ion chamber is then exposed to the beam for a period sufficient to accumulate a dose of several centigray. While this dose is being delivered to the thimble ion chamber, the beam simultaneously produces ionization in the large planar ion chamber upstream of the sample location. Both ion chambers are read out via the NSRL recycling integrator, which digitizes the collected charge in units of counts, with each count representing 10 pC of integrated charge. The dose delivered is calculated by multiplying the integrated charge in the thimble ion chamber by the calibration factor, and this dose is then transferred to the planar ion chamber by associating the delivered dose with the total ionization measured. This calibration remains accurate as long as the beam shape is constant. The nominal dose and dose delivered agree within 10%.

### 3.2.4 Beam fluence

The conversion from (ion chamber) delivered dose to incident particle fluence is performed using the SRIM computational tool (*5*), which calculates the energy deposited by ions of a given energy as they traverse matter. For example, SRIM predicts that a 1000 MeV proton deposits approximately 0.22 keV per micron of path length in water. Since a dose of 1 Gray in water corresponds to the deposition of 1 Joule of energy per kilogram, and 1 J = 6.24E15 keV, this relationship can be used to convert dose to fluence. Considering a 1 cm × 1 cm × 1000 cm water sample with a mass of 1 kg and a total path length of 1E7 microns, a 1000 MeV proton deposits approximately 2.2E6 keV, equivalent to 3.53E-10 J. Therefore, a fluence of 2.84E9 protons/cm² at 1000 MeV is required to deliver 1 Joule of deposited energy, corresponding to a dose of 1 Gray in the water sample.

### 3.2.5 Uncertainties

The primary source of uncertainty in the delivered dose arises from nonuniformity in the beam profile. These nonuniformities are quantified on a spill-by-spill basis by comparing the beam intensity in one region of the beam spot with that in other regions. For radiobiology beams, as used in this work, the root-mean-square (RMS) uniformity is typically maintained within 3%. A portion of this nonuniformity results from spill-to-spill variation, which tends to average out over the course of a multi-spill exposure, yielding overall uniformity better than 3%. Nonuniformity associated with digitization has been determined to be below 1%. Errors in dose delivery due to beam cut-off at the end of an exposure depend on the dose rate but are generally well below 1% for exposures lasting ten or more spills.

## 3.3 SMALL-ANIMAL GCR IRRADIATION

In the experimental setup of this study, the simplified GCR simulation 5-ion beam was utilized to deliver nominal radiation doses of 0.2, 0.4, and 0.75 Gy to adult Balb/c mice (Table 2). The beam was uniform in cross section over the area of occupancy of the mice, to within 3% (*vide supra*). Throughout the irradiation, the fluences of each beam component were monitored and recorded at 6s time intervals. A schematic of the irradiation facility is provided in Fig. 3.

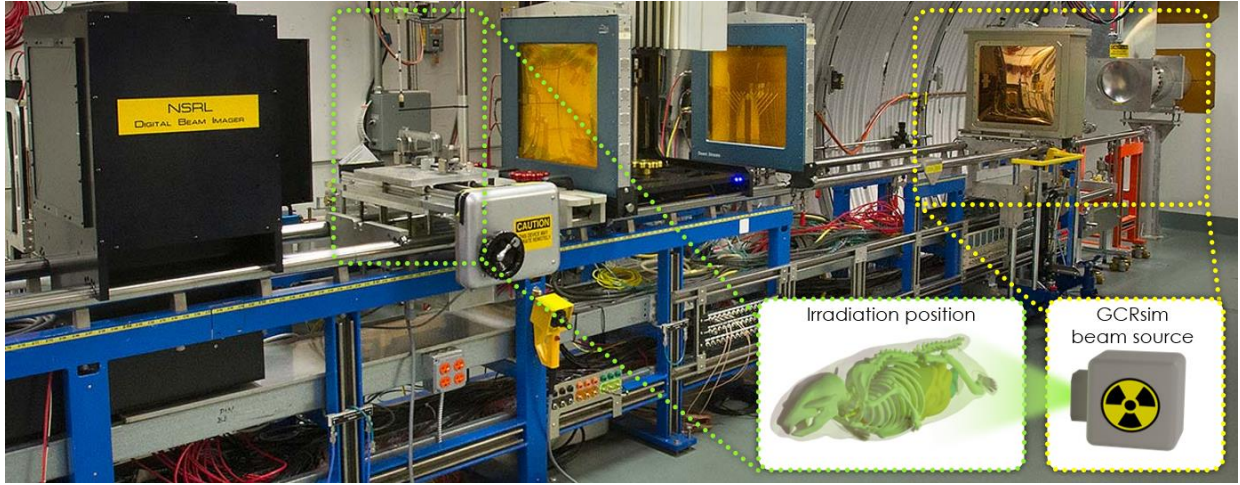

**Figure 3:** Schematic of the GCRsim experimental irradiation setup at BNL. **Figure 3:** Schematic of the GCRsim experimental irradiation setup at BNL. (Photo reproduced with permission from BNL)

All animal studies were conducted under a protocol approved by the Institutional Animal Care and Use Committees (IACUC) of UT Southwestern and Brookhaven National Laboratory.

**Table 2:** Distribution of ion species, their energies, fractions, and corresponding absorbed doses measured via ion chamber, in SimGCRsim irradiations in this work.

| Ion | Energy (MeV/n) | LET (KeV/μm) | Beam Fraction | Dose Delivered (Gy) | | |
|---|---|---|---|---|---|---|
| H | 1000 | 0.22 | 0.35 | 0.07 | 0.14 | 0.263 |
| Si | 600 | 50.24 | 0.01 | 0.002 | 0.004 | 0.0075 |
| He | 250 | 1.58 | 0.18 | 0.036 | 0.072 | 0.135 |
| O | 350 | 20.72 | 0.06 | 0.012 | 0.024 | 0.045 |
| Fe | 600 | 170.01 | 0.01 | 0.002 | 0.004 | 0.0075 |
| H | 250 | 0.40 | 0.39 | 0.078 | 0.156 | 0.293 |
| **Total dose delivered (Gy):** | | | | **0.2** | **0.4** | **0.75** |

### 3.4 MONTE CARLO DOSE RECONSTRUCTION

A core objective of this manuscript is to introduce and validate fluence-to-absorbed dose conversion factors (DCFs) tailored for mouse organs under varying beam orientations. These DCFs are expressed in units of absorbed dose per unit particle fluence, allowing for the calculation of organ-specific dose (or dose rate) by multiplying the measured fluence (or flux) by the DCF for each GCR or GCRsim beam component. If the animal subject is mobile such that the relative beam orientation changes during the irradiation, we consider the following time-dependent absorbed dose formulation:

$$D(r_T) = \int_{t_0}^{t_f} \dot{D}(r_T, t)\, dt \qquad \text{Eqn. 1}$$

with:

$$\dot{D}(r_T, t) = \sum_i \sum_E DCF_{i,E,O(t)}(r_T) \times \phi_{i,E}(t) \qquad \text{Eqn. 2}$$

Where $t$ is the time [s], $D(r_T)$ is the absorbed dose [Gy] to target region $r_T$ accumulated over the irradiation interval $t_0$ to $t_f$, and $\dot{D}(r_T, t)$ is the dose rate [Gy·s⁻¹]. The flux of ion species $i$ with incident energy $E$ [MeV/n] is denoted by $\phi_{i,E}(t)$ [ions·m⁻²·s⁻¹]. The fluence-to-absorbed dose conversion factor is denoted $DCF_{i,E,O(t)}(r_T)$ [Gy·m²·ion⁻¹] where $O(t)$ represents the orientation of the irradiation (e.g., left-right, anterior-posterior, inferior-superior).

If the orientation is fixed (i.e., animal movement is negligible over the exposure interval), then Eqn's 1-2 simplify to:

$$D(r_T) = \sum_i \sum_E DCF_{i,E,O}(r_T) \times \Phi_{i,E} \qquad \text{Eqn. 3}$$

where $\Phi_{i,E}$ [ions·m⁻²] is the ion fluence of species $i$ with incident energy $E$. Of note, the target region of interest may be an organ, a combination of organs (e.g., the total body), or a voxel. For the 'total body' target region, the DCF is defined as a mass-weighted sum of 'living' tissues (e.g., excluding urine, contents of the GI tract, air within the body):

$$DCF_{i,E,O}(total\ body) = \frac{\sum_{r_T \in living\ tissue} DCF_{i,E,O}(r_T) \cdot M(r_T)}{\sum_{r_T \in living\ tissue} M(r_T)} \qquad \text{Eqn. 4}$$

Where $M(r_T)$ is the mass of the target region.

To compute these DCFs, Monte Carlo particle transport simulations were performed in a murine computational anatomic phantom, providing rigorous estimates of murine organ-level and voxel-level radiation doses based on beam orientation. The simulations facilitated the development of a robust model correlating measured fluence to absorbed doses within individual tissues.

Secondarily, we provide conversion factors relating the SimGCRsim ion chamber dose readout (i.e., the delivered dose) to organ-level absorbed dose in murine subjects.

### 3.4.1 Computational Phantom

The MOBY phantom is a well-recognized computational phantom that models gross anatomic structures of a representative mouse, and is widely used for Monte Carlo dosimetry simulations, particularly in the fields of medical imaging and cancer radiation therapy (*6*). This phantom facilitates dosimetry simulations through its detailed representation of mouse anatomy and defines both the spatial boundaries and material compositions within simulations. The phantom was previously modified to tetrahedral mesh format, which offers advantages in computational efficiency, anatomic fidelity when modeling small-scale and detailed anatomical structures, and avoidance of voxelization issues such as stairstep artifacts and partial volume effects (*7,8*). Details of these modifications are described in (*9*).

The phantom was rigidly scaled to a total body mass of 25 g to align with the mean total body mass of the experimental Balb/c rodent model (*10*) used in our SimGCRsim irradiations. Details of the phantom are provided in supplemental data.

### 3.4.2 PHITS Monte Carlo Simulation

Computational simulations of the experimental setup were performed using the 3D Particle and Heavy Ion Transport code System (PHITS), version 3.32(*11*). For the execution of dose computation in PHITS, a coded subroutine was used to incorporate the specific parameters of the irradiation experiment conducted at BNL. The GCR beams were modeled using the PHITS rectangular solid distribution source (s-type = 2) with monoenergetic projectiles given in Table 1. The material outside the phantom was defined as void. The T-Deposit tally was used for computing absorbed dose at the region and voxel level.

In order to simulate the nuclear and atomic collisions, PHITS uses various physics models and data libraries to cover all energy ranges and particle types. In this study, an updated JAERI Quantum Molecular Dynamics model (JQMD 2.0) and/or the intranuclear cascade model (INCL 4.6) was used to simulate the nucleus-nucleus reactions in high-energy collisions (*12–14*). The Generalized Evaporation Model (GEM) algorithm was also implemented to simulate the de-excitation process (*15*). The high energy Electron Gamma Shower (EGS5) mode was used for transport of photons, electrons and positrons (*16*).

At least $10^7$ histories were simulated for each source particle, which maintained the statistical error margin for the mean organ absorbed doses below 1% for all organs of the phantom.

# 4 RESULTS AND DISCUSSION

## 4.1 ORGAN-LEVEL FLUENCE-TO-ABSORBED DOSE CONVERSION FACTORS FOR MURINE SIMGCRSIM IRRADIATIONS

Tables 3-8 summarize the simulated mouse DCFs for the various components of the SimGCRsim under different beam orientations.

**Table 3:** Fluence-to-absorbed dose conversion factors for 1000 MeV protons.

| | $DCF_{H,1000\ MeV/n}$ [Gy·cm²·ion⁻¹] | | | | | |
|---|---|---|---|---|---|---|
| Beam orientation | L-R | R-L | A-P | P-A | I-S | S-I |
| Esophagus | 4.37E-10 | 4.28E-10 | 4.23E-10 | 4.30E-10 | 4.20E-10 | 4.22E-10 |
| Stomach wall | 4.22E-10 | 4.40E-10 | 4.35E-10 | 4.29E-10 | 4.29E-10 | 4.28E-10 |
| Small intestine wall | 4.36E-10 | 4.29E-10 | 4.29E-10 | 4.37E-10 | 4.31E-10 | 4.28E-10 |
| Large intestine wall | 4.34E-10 | 4.37E-10 | 4.33E-10 | 4.36E-10 | 4.32E-10 | 4.26E-10 |
| Gallbladder | 4.24E-10 | 4.30E-10 | 4.26E-10 | 4.43E-10 | 4.31E-10 | 4.29E-10 |
| Liver | 4.32E-10 | 4.30E-10 | 4.30E-10 | 4.35E-10 | 4.29E-10 | 4.27E-10 |
| Pancreas | 4.30E-10 | 4.40E-10 | 4.33E-10 | 4.36E-10 | 4.32E-10 | 4.29E-10 |
| Heart wall | 4.33E-10 | 4.32E-10 | 4.26E-10 | 4.35E-10 | 4.30E-10 | 4.27E-10 |
| Lungs | 4.28E-10 | 4.27E-10 | 4.29E-10 | 4.29E-10 | 4.26E-10 | 4.24E-10 |
| Cortical bone | 3.87E-10 | 3.87E-10 | 3.90E-10 | 3.86E-10 | 3.82E-10 | 3.83E-10 |
| Bone marrow | 4.30E-10 | 4.30E-10 | 4.31E-10 | 4.28E-10 | 4.21E-10 | 4.24E-10 |
| Kidneys | 4.36E-10 | 4.34E-10 | 4.38E-10 | 4.29E-10 | 4.32E-10 | 4.27E-10 |
| Urinary bladder wall | 4.42E-10 | 4.35E-10 | 4.27E-10 | 4.38E-10 | 4.31E-10 | 4.21E-10 |
| Testes | 4.30E-10 | 4.28E-10 | 4.24E-10 | 4.30E-10 | 4.22E-10 | 4.17E-10 |
| Brain | 4.22E-10 | 4.26E-10 | 4.32E-10 | 4.21E-10 | 4.16E-10 | 4.22E-10 |
| Skin | 4.00E-10 | 4.00E-10 | 4.00E-10 | 4.00E-10 | 3.97E-10 | 3.99E-10 |
| Spleen | 4.20E-10 | 4.38E-10 | 4.31E-10 | 4.22E-10 | 4.23E-10 | 4.24E-10 |
| Thyroid | 4.42E-10 | 4.32E-10 | 4.28E-10 | 4.38E-10 | 4.26E-10 | 4.25E-10 |
| Residual soft tissue | 4.26E-10 | 4.26E-10 | 4.27E-10 | 4.26E-10 | 4.22E-10 | 4.22E-10 |
| Total body | 4.23E-10 | 4.23E-10 | 4.24E-10 | 4.24E-10 | 4.19E-10 | 4.19E-10 |

**Table 4:** Fluence-to-absorbed dose conversion factors for 250 MeV protons.

| | $DCF_{H,250\ MeV/n}$ [Gy·cm$^2$·ion$^{-1}$] | | | | | |
|---|---|---|---|---|---|---|
| Beam orientation | L-R | R-L | A-P | P-A | I-S | S-I |
| Esophagus | 6.81E-10 | 6.82E-10 | 6.76E-10 | 6.86E-10 | 6.89E-10 | 6.78E-10 |
| Stomach wall | 6.77E-10 | 6.96E-10 | 6.88E-10 | 6.82E-10 | 6.90E-10 | 6.88E-10 |
| Small intestine wall | 6.92E-10 | 6.83E-10 | 6.82E-10 | 6.91E-10 | 6.89E-10 | 6.90E-10 |
| Large intestine wall | 6.85E-10 | 6.92E-10 | 6.87E-10 | 6.90E-10 | 6.89E-10 | 6.90E-10 |
| Gallbladder | 6.91E-10 | 6.92E-10 | 6.82E-10 | 7.06E-10 | 6.94E-10 | 6.86E-10 |
| Liver | 6.86E-10 | 6.83E-10 | 6.82E-10 | 6.88E-10 | 6.89E-10 | 6.85E-10 |
| Pancreas | 6.85E-10 | 6.97E-10 | 6.87E-10 | 6.90E-10 | 6.92E-10 | 6.90E-10 |
| Heart wall | 6.81E-10 | 6.86E-10 | 6.79E-10 | 6.90E-10 | 6.93E-10 | 6.85E-10 |
| Lungs | 6.84E-10 | 6.81E-10 | 6.81E-10 | 6.81E-10 | 6.88E-10 | 6.81E-10 |
| Cortical bone | 6.13E-10 | 6.13E-10 | 6.16E-10 | 6.11E-10 | 6.15E-10 | 6.13E-10 |
| Bone marrow | 6.95E-10 | 6.95E-10 | 6.97E-10 | 6.94E-10 | 6.99E-10 | 6.94E-10 |
| Kidneys | 6.86E-10 | 6.87E-10 | 6.92E-10 | 6.80E-10 | 6.90E-10 | 6.88E-10 |
| Urinary bladder wall | 6.87E-10 | 6.87E-10 | 6.79E-10 | 6.93E-10 | 6.88E-10 | 6.90E-10 |
| Testes | 6.84E-10 | 6.83E-10 | 6.77E-10 | 6.86E-10 | 6.76E-10 | 6.89E-10 |
| Brain | 6.79E-10 | 6.82E-10 | 6.87E-10 | 6.75E-10 | 6.87E-10 | 6.79E-10 |
| Skin | 6.64E-10 | 6.63E-10 | 6.63E-10 | 6.63E-10 | 6.64E-10 | 6.65E-10 |
| Spleen | 6.70E-10 | 6.93E-10 | 6.87E-10 | 6.72E-10 | 6.83E-10 | 6.84E-10 |
| Thyroid | 6.84E-10 | 6.81E-10 | 6.78E-10 | 6.88E-10 | 6.92E-10 | 6.81E-10 |
| Residual soft tissue | 6.91E-10 | 6.91E-10 | 6.91E-10 | 6.90E-10 | 6.92E-10 | 6.92E-10 |
| Total body | 6.83E-10 | 6.83E-10 | 6.83E-10 | 6.82E-10 | 6.84E-10 | 6.84E-10 |

**Table 5:** Fluence-to-absorbed dose conversion factors for 250 MeV/n helium ions.

| | $DCF_{He,250\ MeV/n}$ [Gy·cm$^2$·ion$^{-1}$] | | | | | |
|---|---|---|---|---|---|---|
| Beam orientation | L-R | R-L | A-P | P-A | I-S | S-I |
| Esophagus | 2.62E-09 | 2.65E-09 | 2.60E-09 | 2.61E-09 | 2.62E-09 | 2.61E-09 |
| Stomach wall | 2.66E-09 | 2.64E-09 | 2.64E-09 | 2.63E-09 | 2.65E-09 | 2.64E-09 |
| Small intestine wall | 2.63E-09 | 2.62E-09 | 2.63E-09 | 2.65E-09 | 2.64E-09 | 2.64E-09 |
| Large intestine wall | 2.65E-09 | 2.64E-09 | 2.64E-09 | 2.65E-09 | 2.66E-09 | 2.64E-09 |
| Gallbladder | 2.65E-09 | 2.60E-09 | 2.66E-09 | 2.69E-09 | 2.68E-09 | 2.67E-09 |
| Liver | 2.63E-09 | 2.63E-09 | 2.63E-09 | 2.64E-09 | 2.65E-09 | 2.63E-09 |
| Pancreas | 2.66E-09 | 2.65E-09 | 2.64E-09 | 2.64E-09 | 2.66E-09 | 2.64E-09 |
| Heart wall | 2.64E-09 | 2.66E-09 | 2.63E-09 | 2.65E-09 | 2.66E-09 | 2.64E-09 |
| Lungs | 2.62E-09 | 2.63E-09 | 2.63E-09 | 2.63E-09 | 2.64E-09 | 2.62E-09 |
| Cortical bone | 2.36E-09 | 2.37E-09 | 2.37E-09 | 2.35E-09 | 2.36E-09 | 2.36E-09 |
| Bone marrow | 2.68E-09 | 2.69E-09 | 2.68E-09 | 2.68E-09 | 2.68E-09 | 2.67E-09 |
| Kidneys | 2.64E-09 | 2.68E-09 | 2.66E-09 | 2.62E-09 | 2.65E-09 | 2.64E-09 |
| Urinary bladder wall | 2.64E-09 | 2.68E-09 | 2.63E-09 | 2.68E-09 | 2.65E-09 | 2.63E-09 |
| Testes | 2.63E-09 | 2.67E-09 | 2.61E-09 | 2.63E-09 | 2.62E-09 | 2.63E-09 |
| Brain | 2.63E-09 | 2.63E-09 | 2.64E-09 | 2.60E-09 | 2.64E-09 | 2.63E-09 |
| Skin | 2.58E-09 | 2.57E-09 | 2.58E-09 | 2.58E-09 | 2.58E-09 | 2.58E-09 |
| Spleen | 2.66E-09 | 2.68E-09 | 2.63E-09 | 2.61E-09 | 2.62E-09 | 2.64E-09 |
| Thyroid | 2.62E-09 | 2.60E-09 | 2.62E-09 | 2.63E-09 | 2.63E-09 | 2.61E-09 |
| Residual soft tissue | 2.66E-09 | 2.66E-09 | 2.66E-09 | 2.66E-09 | 2.66E-09 | 2.66E-09 |
| Total body | 2.63E-09 | 2.63E-09 | 2.63E-09 | 2.63E-09 | 2.63E-09 | 2.63E-09 |

**Table 6:** Fluence-to-absorbed dose conversion factors for 350 MeV/n oxygen ions.

| | $DCF_{\text{O,350 MeV/n}}$ [Gy·cm$^2$·ion$^{-1}$] | | | | | |
|---|---|---|---|---|---|---|
| Beam orientation | L-R | R-L | A-P | P-A | I-S | S-I |
| Esophagus | 3.33E-08 | 3.40E-08 | 3.33E-08 | 3.31E-08 | 3.27E-08 | 3.29E-08 |
| Stomach wall | 3.34E-08 | 3.30E-08 | 3.34E-08 | 3.34E-08 | 3.32E-08 | 3.33E-08 |
| Small intestine wall | 3.35E-08 | 3.35E-08 | 3.35E-08 | 3.35E-08 | 3.32E-08 | 3.32E-08 |
| Large intestine wall | 3.35E-08 | 3.33E-08 | 3.36E-08 | 3.35E-08 | 3.34E-08 | 3.33E-08 |
| Gallbladder | 3.33E-08 | 3.34E-08 | 3.42E-08 | 3.37E-08 | 3.32E-08 | 3.33E-08 |
| Liver | 3.35E-08 | 3.34E-08 | 3.34E-08 | 3.33E-08 | 3.30E-08 | 3.31E-08 |
| Pancreas | 3.36E-08 | 3.34E-08 | 3.35E-08 | 3.33E-08 | 3.32E-08 | 3.32E-08 |
| Heart wall | 3.33E-08 | 3.37E-08 | 3.36E-08 | 3.36E-08 | 3.32E-08 | 3.33E-08 |
| Lungs | 3.34E-08 | 3.34E-08 | 3.35E-08 | 3.34E-08 | 3.30E-08 | 3.31E-08 |
| Cortical bone | 3.01E-08 | 3.01E-08 | 3.00E-08 | 3.00E-08 | 2.99E-08 | 2.99E-08 |
| Bone marrow | 3.43E-08 | 3.41E-08 | 3.42E-08 | 3.42E-08 | 3.43E-08 | 3.39E-08 |
| Kidneys | 3.36E-08 | 3.39E-08 | 3.34E-08 | 3.35E-08 | 3.31E-08 | 3.31E-08 |
| Urinary bladder wall | 3.35E-08 | 3.36E-08 | 3.35E-08 | 3.35E-08 | 3.35E-08 | 3.33E-08 |
| Testes | 3.35E-08 | 3.39E-08 | 3.34E-08 | 3.32E-08 | 3.36E-08 | 3.35E-08 |
| Brain | 3.37E-08 | 3.36E-08 | 3.35E-08 | 3.36E-08 | 3.36E-08 | 3.36E-08 |
| Skin | 3.33E-08 | 3.33E-08 | 3.33E-08 | 3.34E-08 | 3.32E-08 | 3.32E-08 |
| Spleen | 3.35E-08 | 3.37E-08 | 3.32E-08 | 3.36E-08 | 3.33E-08 | 3.32E-08 |
| Thyroid | 3.34E-08 | 3.27E-08 | 3.34E-08 | 3.30E-08 | 3.30E-08 | 3.31E-08 |
| Residual soft tissue | 3.39E-08 | 3.39E-08 | 3.39E-08 | 3.39E-08 | 3.37E-08 | 3.37E-08 |
| Total body | 3.35E-08 | 3.35E-08 | 3.35E-08 | 3.35E-08 | 3.33E-08 | 3.33E-08 |

**Table 7:** Fluence-to-absorbed dose conversion factors for 600 MeV/n silicon ions.

| | $DCF_{\text{Si,600 MeV/n}}$ [Gy·cm$^2$·ion$^{-1}$] | | | | | |
|---|---|---|---|---|---|---|
| Beam orientation | L-R | R-L | A-P | P-A | I-S | S-I |
| Esophagus | 7.92E-08 | 8.11E-08 | 7.94E-08 | 7.90E-08 | 7.04E-08 | 7.57E-08 |
| Stomach wall | 8.05E-08 | 7.69E-08 | 7.90E-08 | 7.97E-08 | 7.49E-08 | 7.62E-08 |
| Small intestine wall | 7.89E-08 | 7.97E-08 | 8.01E-08 | 7.92E-08 | 7.66E-08 | 7.43E-08 |
| Large intestine wall | 7.98E-08 | 7.84E-08 | 7.99E-08 | 7.91E-08 | 7.74E-08 | 7.30E-08 |
| Gallbladder | 7.91E-08 | 7.93E-08 | 8.19E-08 | 7.81E-08 | 7.49E-08 | 7.69E-08 |
| Liver | 7.91E-08 | 7.93E-08 | 7.96E-08 | 7.88E-08 | 7.43E-08 | 7.59E-08 |
| Pancreas | 8.02E-08 | 7.82E-08 | 7.97E-08 | 7.88E-08 | 7.54E-08 | 7.44E-08 |
| Heart wall | 7.93E-08 | 8.02E-08 | 8.01E-08 | 7.88E-08 | 7.33E-08 | 7.63E-08 |
| Lungs | 7.94E-08 | 7.98E-08 | 7.98E-08 | 7.97E-08 | 7.30E-08 | 7.60E-08 |
| Cortical bone | 7.19E-08 | 7.19E-08 | 7.15E-08 | 7.20E-08 | 6.68E-08 | 6.87E-08 |
| Bone marrow | 8.15E-08 | 8.11E-08 | 8.12E-08 | 8.19E-08 | 7.44E-08 | 7.85E-08 |
| Kidneys | 7.98E-08 | 8.02E-08 | 7.87E-08 | 7.99E-08 | 7.59E-08 | 7.39E-08 |
| Urinary bladder wall | 7.98E-08 | 7.95E-08 | 8.04E-08 | 7.91E-08 | 7.91E-08 | 7.11E-08 |
| Testes | 7.94E-08 | 8.02E-08 | 8.03E-08 | 7.87E-08 | 8.03E-08 | 7.01E-08 |
| Brain | 8.04E-08 | 8.01E-08 | 7.95E-08 | 8.06E-08 | 7.09E-08 | 7.98E-08 |
| Skin | 7.98E-08 | 7.96E-08 | 7.99E-08 | 8.00E-08 | 7.73E-08 | 7.76E-08 |
| Spleen | 8.07E-08 | 7.86E-08 | 7.85E-08 | 8.07E-08 | 7.63E-08 | 7.71E-08 |
| Thyroid | 7.99E-08 | 7.83E-08 | 7.98E-08 | 7.79E-08 | 7.11E-08 | 7.65E-08 |
| Residual soft tissue | 8.05E-08 | 8.05E-08 | 8.07E-08 | 8.07E-08 | 7.68E-08 | 7.66E-08 |
| Total body | 7.96E-08 | 7.96E-08 | 7.97E-08 | 7.97E-08 | 7.56E-08 | 7.58E-08 |

**Table 8:** Fluence-to-absorbed dose conversion factors for 600 MeV/n iron ions.

| | $DCF_{\text{Fe,600 MeV/n}}$ [Gy·cm$^2$·ion$^{-1}$] | | | | | |
|---|---|---|---|---|---|---|
| Beam orientation | L-R | R-L | A-P | P-A | I-S | S-I |
| Esophagus | 2.80E-07 | 2.78E-07 | 2.77E-07 | 2.72E-07 | 2.65E-07 | 2.65E-07 |
| Stomach wall | 2.79E-07 | 2.74E-07 | 2.77E-07 | 2.76E-07 | 2.69E-07 | 2.71E-07 |
| Small intestine wall | 2.77E-07 | 2.79E-07 | 2.80E-07 | 2.77E-07 | 2.72E-07 | 2.69E-07 |
| Large intestine wall | 2.78E-07 | 2.76E-07 | 2.79E-07 | 2.77E-07 | 2.73E-07 | 2.68E-07 |
| Gallbladder | 2.77E-07 | 2.77E-07 | 2.83E-07 | 2.79E-07 | 2.69E-07 | 2.74E-07 |
| Liver | 2.77E-07 | 2.78E-07 | 2.78E-07 | 2.76E-07 | 2.67E-07 | 2.70E-07 |
| Pancreas | 2.80E-07 | 2.76E-07 | 2.77E-07 | 2.77E-07 | 2.69E-07 | 2.68E-07 |
| Heart wall | 2.77E-07 | 2.77E-07 | 2.80E-07 | 2.73E-07 | 2.69E-07 | 2.73E-07 |
| Lungs | 2.79E-07 | 2.79E-07 | 2.78E-07 | 2.78E-07 | 2.66E-07 | 2.70E-07 |
| Cortical bone | 2.52E-07 | 2.51E-07 | 2.50E-07 | 2.52E-07 | 2.45E-07 | 2.47E-07 |
| Bone marrow | 2.84E-07 | 2.85E-07 | 2.84E-07 | 2.87E-07 | 2.79E-07 | 2.78E-07 |
| Kidneys | 2.80E-07 | 2.78E-07 | 2.76E-07 | 2.78E-07 | 2.71E-07 | 2.68E-07 |
| Urinary bladder wall | 2.76E-07 | 2.78E-07 | 2.81E-07 | 2.79E-07 | 2.77E-07 | 2.69E-07 |
| Testes | 2.79E-07 | 2.78E-07 | 2.79E-07 | 2.79E-07 | 2.82E-07 | 2.73E-07 |
| Brain | 2.81E-07 | 2.79E-07 | 2.78E-07 | 2.83E-07 | 2.70E-07 | 2.79E-07 |
| Skin | 2.79E-07 | 2.79E-07 | 2.79E-07 | 2.79E-07 | 2.76E-07 | 2.76E-07 |
| Spleen | 2.83E-07 | 2.73E-07 | 2.78E-07 | 2.81E-07 | 2.71E-07 | 2.72E-07 |
| Thyroid | 2.80E-07 | 2.77E-07 | 2.76E-07 | 2.71E-07 | 2.65E-07 | 2.71E-07 |
| Residual soft tissue | 2.82E-07 | 2.81E-07 | 2.82E-07 | 2.82E-07 | 2.75E-07 | 2.75E-07 |
| Total body | 2.78E-07 | 2.78E-07 | 2.79E-07 | 2.79E-07 | 2.72E-07 | 2.72E-07 |

## 4.2 DELIVERED DOSE-TO-ORGAN ABSORBED DOSE CONVERSION FACTORS FOR MURINE SIMGCRSIM IRRADIATIONS

Table 9 provides conversion factors relating the SimGCRsim nominal dose or ion chamber dose readout (i.e., the delivered dose) to organ-level absorbed dose in murine subjects. These conversion factors will be valid for the beam energies and beam fractions described in Table 2.

**Table 9:** Delivered dose-to-mouse organ absorbed dose conversion factors for the SimGCRsim beam.

| | Conversion Factor [Gy/Gy$_{delivered}$] | | | | | |
|---|---|---|---|---|---|---|
| Beam orientation | L-R | R-L | A-P | P-A | I-S | S-I |
| Esophagus | 1.12 | 1.11 | 1.10 | 1.11 | 1.10 | 1.10 |
| Stomach wall | 1.10 | 1.13 | 1.12 | 1.11 | 1.11 | 1.11 |
| Small intestine wall | 1.12 | 1.11 | 1.11 | 1.12 | 1.11 | 1.11 |
| Large intestine wall | 1.12 | 1.12 | 1.12 | 1.12 | 1.12 | 1.11 |
| Gallbladder | 1.11 | 1.11 | 1.11 | 1.14 | 1.12 | 1.11 |
| Liver | 1.11 | 1.11 | 1.11 | 1.12 | 1.11 | 1.11 |
| Pancreas | 1.11 | 1.13 | 1.12 | 1.12 | 1.12 | 1.11 |
| Heart wall | 1.11 | 1.12 | 1.10 | 1.12 | 1.12 | 1.11 |
| Lungs | 1.11 | 1.11 | 1.11 | 1.11 | 1.11 | 1.10 |
| Cortical bone | 1.00 | 1.00 | 1.00 | 0.99 | 0.99 | 0.99 |
| Bone marrow | 1.12 | 1.12 | 1.12 | 1.12 | 1.12 | 1.11 |
| Kidneys | 1.12 | 1.12 | 1.13 | 1.11 | 1.12 | 1.11 |
| Urinary bladder wall | 1.13 | 1.12 | 1.10 | 1.13 | 1.12 | 1.10 |
| Testes | 1.11 | 1.11 | 1.10 | 1.11 | 1.10 | 1.10 |
| Brain | 1.10 | 1.11 | 1.12 | 1.09 | 1.10 | 1.10 |
| Skin | 1.06 | 1.06 | 1.06 | 1.06 | 1.06 | 1.06 |
| Spleen | 1.10 | 1.13 | 1.11 | 1.09 | 1.10 | 1.11 |
| Thyroid | 1.12 | 1.11 | 1.10 | 1.12 | 1.11 | 1.10 |
| Residual soft tissue | 1.11 | 1.11 | 1.11 | 1.11 | 1.11 | 1.11 |
| Total body | 1.10 | 1.10 | 1.10 | 1.10 | 1.10 | 1.10 |

## 4.3 VOXEL-LEVEL FLUENCE-TO-ABSORBED DOSE CONVERSION FACTORS FOR MURINE SIMGCRSIM IRRADIATIONS

Voxel-level DCFs for the various components and orientations of the SimGCRsim are available in the electronic supplemental data files.

## 4.4 METHOD VALIDATION AND BEAM CALIBRATION ASSESSMENT

Validation of our DCF-based method was conducted through experimental irradiations of mice at multiple nominal/delivered dose levels (0.20, 0.40, and 0.75 Gy). This process tested the accuracy of the DCFs by comparing the nominal (delivered) absorbed doses with absorbed dose to murine tissues calculated from the measured fluence and corresponding dose conversion factors. This addresses the linearity and reliability of our method across a spectrum of absorbed doses.

In the GCRsim configuration, input irradiation parameters allow for the specification of a nominal radiation dose which is calibrated to ion chamber measurement in reference geometry. However, due to the lack of specific calibrations for subjects with irregular geometries such as mice, the reported delivered dose may differ substantially from the actual absorbed dose to the subject. Our developed library of DCFs compensates for these inaccuracies by adjusting the doses based on the specific radiation profiles of different GCR species. Analysis of the relationship between nominal doses and those calculated from beam fluence using our DCFs shows a linear correlation (Figure 4), enabling the derivation of a 'calibration factor' or correction factor. This factor aligns the prescribed radiation doses with the actual doses received by the mice, thereby refining the dosimetry process within our experimental framework.

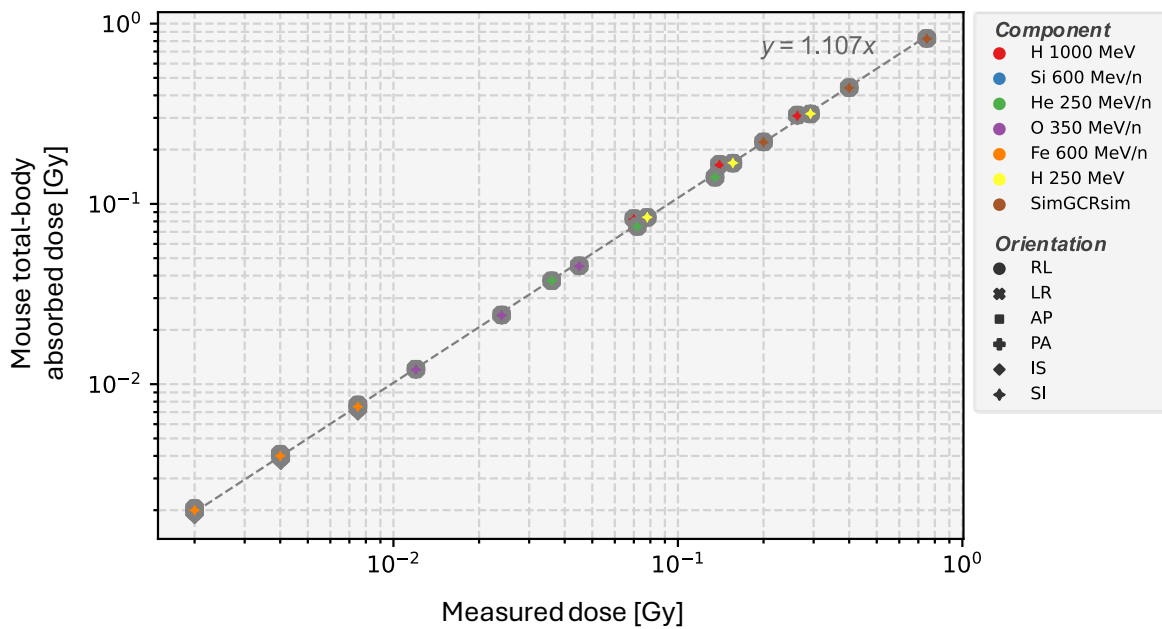

**Figure 4:** Calibration curve relating ion chamber physical dose measurements to total body absorbed dose derived from DCF library and beam fluence measurements.

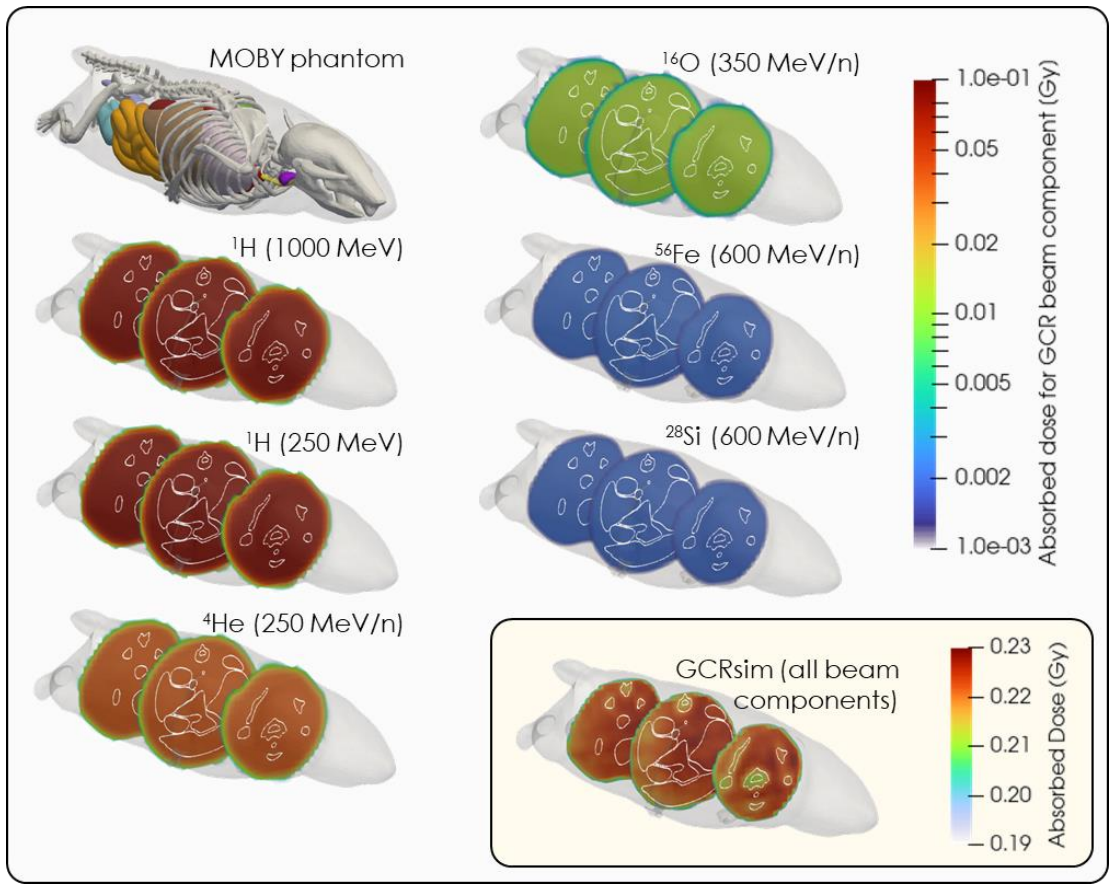

**Figure 5:** Voxel level absorbed dose maps illustrate the relative dose contributions of the various GCRsim ions. *Inset:* Summed dose maps show the spatial variations in absorbed dose (note lower doses received by bone tissue).

Small differences were observed between murine organ-specific doses and the total body dose; these differences can be attributed to the modest attenuation and changes in stopping power as ions traverse the mouse body. This suggests that the dose correction factors (DCFs) generated here are largely robust for small-animal models. However, for larger animal irradiations, the attenuation and variation in stopping power with depth is expected to become more pronounced, potentially leading to greater organ-to-organ dose differentials. A future direction would be to extend this approach to generate DCF libraries for larger animal models, and eventually for humans, to support space radiation bioeffect assessment in these species.

Some limitations should be acknowledged. Relative biological effectiveness (RBE) was not considered in this work; nonetheless, the present dosimetric findings provide a foundation for future efforts to establish RBE values in relevant biological endpoints. Additionally, the phantom model employed does not explicitly incorporate microstructural features such as bone trabeculae which may influence local dose distributions in the skeleton. While these simplifications were necessary for computational feasibility, they represent areas for refinement in future modeling efforts. Finally, our simulations showed that both cortical bone and skin received lower doses relative to internal organs and the total body, as shown in Figure 6. This effect reflects the reduced energy deposition in high-density cortical structures and superficial tissues under the irradiation conditions considered. These findings emphasize the need to account for tissue-specific dose variations when interpreting biological outcomes, particularly in tissues with distinct physical or anatomical characteristics.

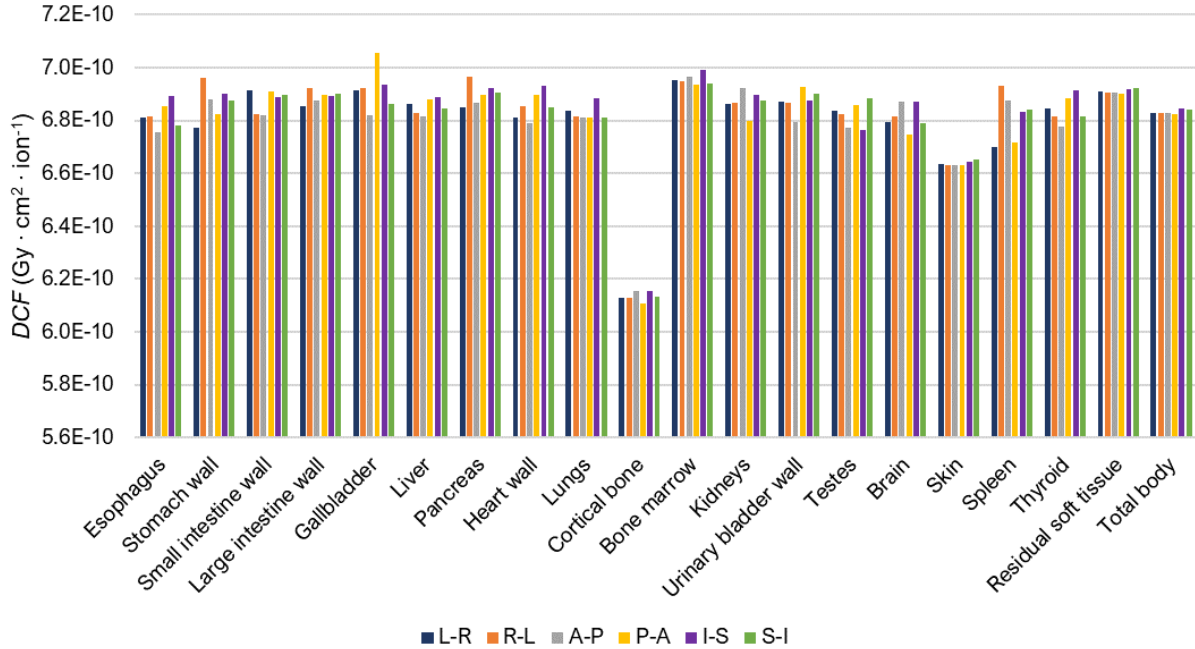

**Figure 6:** Dose conversion factors (DCF, Gy·cm²·ion⁻¹) for individual organs and total body in the mouse phantom under hydrogen 250 MeV/n energy irradiation. Results are shown for six beam orientations: left-to-right (L–R), right-to-left (R–L), anterior-to-posterior (A–P), posterior-to-anterior (P–A), inferior-to-superior (I–S), and superior-to-inferior (S–I).

Future work includes extension of these methods to larger animal models, which are more representative of human biology and therefore more suitable for evaluating the biological impact of cosmic radiation. Additional studies should also utilize the full GCRsim beam rather than the reduced SimGCRsim configuration in order to capture the complete spectrum of radiation conditions encountered in space.

For larger animals, the assumption of a uniform irradiation field may not be valid. In such cases, image-guided dosimetry could be implemented, using beam imager measurements to define the incident ion distribution on the phantom. This approach would allow more accurate modeling of spatially nonuniform dose delivery.

## 5 CONCLUSION

Fluence-to-dose conversion factors were derived to yield accurate murine organ dose measurements that account for variations in stopping power as ions traverse the mouse body. Relatively small differences in organ doses compared with the total body dose were observed. These findings provide a reliable foundation for preclinical studies, although caution is warranted when extrapolating to larger animals, where depth-dependent variations in stopping power may lead to greater organ-to-organ dose differentials. Future work should extend these methods to larger animal models and incorporate full Galactic Cosmic Ray simulations to better represent space radiation conditions. Additionally, improvements in phantom modeling, including microstructural features and image-guided dosimetry, may further enhance the spatial accuracy of organ-specific dose estimates. Overall, these results underscore the significance of accounting for tissue-specific dose differences when evaluating biological outcomes, especially in structures with unique physical or anatomical properties.

# 6 ACKNOWLEDGEMENTS

We greatly acknowledge support from National Aeronautics and Space Agency grant 80NSSC18K1676 (MDS), the David A. Pistenmaa, MD, PhD Distinguished Chair in Radiation Oncology (MDS) and the NIH/NCI Cancer Center Support Grant P30 CA008748 to MSK (LMC). In addition, we thank the NASA Space Radiation Laboratory and the Animal Care personnel at Brookhaven National Laboratory for their help and support.

# 7 DISCLOSURE

LMC serves as a consultant for Evergreen Theragnostics. No other conflict of interest was reported.